\documentclass[aps,prb,superscriptaddress,floatfix]{revtex4}
\usepackage{graphicx,graphics}
\usepackage{dcolumn}
\usepackage{amsmath,amssymb,amsfonts}
\usepackage{latexsym,verbatim}
\usepackage{color}
\usepackage{ulem}
\usepackage{latexsym,verbatim}

\begin{document}

\title{Zero bias conductance peak in detached flakes of superconducting 2H-TaS$_2$ probed by scanning tunneling spectroscopy}

\author{J. A. Galvis} 
\affiliation{Laboratorio de Bajas Temperaturas, Departamento de F\'isica de la Materia Condensada, Instituto de Ciencia de Materiales Nico\'las Cabrera, Condensed Matter Physics Center (IFIMAC), Universidad Aut\'onoma de Madrid, E-28049 Madrid, Spain}
\author{L. Chirolli}
\affiliation{Instituto de Ciencia de Materiales de Madrid (CSIC), Sor Juana In\'ez de la Cruz 3, E-28049 Madrid, Spain}
\author{I. Guillam\'on}
\affiliation{Laboratorio de Bajas Temperaturas, Departamento de F\'isica de la Materia Condensada, Instituto de Ciencia de Materiales Nico\'las Cabrera, Condensed Matter Physics Center (IFIMAC), Universidad Aut\'onoma de Madrid, E-28049 Madrid, Spain}
\affiliation{Unidad Asociada de Bajas Temperaturas y Altos Campos Magn\'eticos, UAM, CSIC, Cantoblanco, E-28049 Madrid, Spain}
\author{S. Vieira}
\affiliation{Laboratorio de Bajas Temperaturas, Departamento de F\'isica de la Materia Condensada, Instituto de Ciencia de Materiales Nico\'las Cabrera, Condensed Matter Physics Center (IFIMAC), Universidad Aut\'onoma de Madrid, E-28049 Madrid, Spain}
\affiliation{Unidad Asociada de Bajas Temperaturas y Altos Campos Magn\'eticos, UAM, CSIC, Cantoblanco, E-28049 Madrid, Spain}
\author{E. Navarro-Moratalla} 
\affiliation{Instituto de Ciencia Molecular (ICMol), Universidad de Valencia, Catedr\'atico Jos\'e Beltr\'an 2, 46980 Paterna, Spain}
\author{E. Coronado} 
\affiliation{Instituto de Ciencia Molecular (ICMol), Universidad de Valencia, Catedr\'atico Jos\'e Beltr\'an 2, 46980 Paterna, Spain}
\author{H. Suderow}
\affiliation{Laboratorio de Bajas Temperaturas, Departamento de F\'isica de la Materia Condensada, Instituto de Ciencia de Materiales Nico\'las Cabrera, Condensed Matter Physics Center (IFIMAC), Universidad Aut\'onoma de Madrid, E-28049 Madrid, Spain}
\affiliation{Unidad Asociada de Bajas Temperaturas y Altos Campos Magn\'eticos, UAM, CSIC, Cantoblanco, E-28049 Madrid, Spain}
\author{F. Guinea}
\affiliation{Instituto de Ciencia de Materiales de Madrid (CSIC), Sor Juana In\'ez de la Cruz 3, E-28049 Madrid, Spain}
\affiliation{Unidad Asociada de Bajas Temperaturas y Altos Campos Magn\'eticos, UAM, CSIC, Cantoblanco, E-28049 Madrid, Spain}

\begin{abstract}
We report an anomalous tunneling conductance with a zero bias peak in flakes of superconducting 2H-TaS$_2$ detached through mechanical exfoliation. To explain the observed phenomenon, we construct a minimal model for a single unit cell layer of superconducting 2H-TaS$_2$ with a simplified 2D Fermi surface and sign-changing Cooper pair wavefunction induced by Coulomb repulsion. Superconductivity is induced in the central $\Gamma$ pocket, where it becomes nodal. We show that weak scattering at the nodal Fermi surface, produced by non-perturbative coupling between tip and sample, gives Andreev states that lead to a zero bias peak in the tunneling conductance. We suggest that reducing dimensionality down to a few atom thick crystals could drive a crossover from conventional to sign changing pairing in the superconductor 2H-TaS$_2$.
\end{abstract}

\maketitle

\section{Introduction.}

The theoretical view of single layer and other quasi-two dimensional superconductors is at odds with the experiments available until now. Theoretical proposals point out that, in 2D, a non-simply connected Fermi surface promotes non-conventional superconductivity mediated by repulsive electron-electron interactions, that favor Kohn-Luttinger related pairing 
\cite{KohnLuttinger1965}. Novel superconducting states include odd momentum pairing in graphene heterostructures \cite{Guinea12}, chiral superconductivity in doped graphene \cite{Chubukov12} or sign changing superconductivity in MoS$_2$ \cite{Roldan13}. Superconductivity has  been found in different 2D systems, like single atom layers  
\cite{Qin09,Wang12,Gardner11,Tominaga13,Cren09}, unit cell layers of heavy fermion compounds and in iron selenium \cite{Mizukami11,Liu12,Hirschfeld11}, electrons confined at interfaces \cite{Ahn03,Bollinger11}, or in electron-doped MoS$_2$\cite{Taniguchi12,Ye12}.  Experiments have studied phase diagrams as a function of carrier density, magnetic field and temperature\cite{Qin09,Liu12,Wang12,Gardner11,Tominaga13,Cren09,Mizukami11,Ahn03,Bollinger11,Taniguchi12,Ye12,Hirschfeld11}. Some tunneling experiments have been made \cite{Tominaga13,Cren09,Liu12}, reporting conventional single or multigap s-wave like superconducting tunneling features, without clear indications for Kohn-Luttinger related pairing.

Andreev bound states (ABS) are formed by scattering with defects or non-magnetic impurities in Cooper pair wavefunction sign-changing non-conventional superconductors \cite{vanWessPRL1992} and manifest as a peak in the tunneling conductance, often at zero bias (zero bias conductance peak, ZBCP). ABS are not formed in conventional s-wave BCS superconductors with defects or non-magnetic impurities 
\cite{Tsuei00,Balatsky06,Hirschfeld11,Hoffman11,Wang12,HuPRL1994,TanakaPRL1995,KashiwayaPRB1995,HuPRB1998,
KashiwayaRPP2000,Bascones01,SatoPRB2011,DahlhausPRB2012}. ABS are useful to produce the conditions favoring topological superconductivity and Majorana fermion physics \cite{Qi-Zhang,Beenakker}, because they give a large amount of states at the Fermi level, often with linear dispersion relation\cite{Balatsky06}. 

Here we present Scanning Tunneling Microscopy (STM) measurements in flakes detached from the surface of the transition metal dichalcogenide 2H-TaS$_2$. We find a ZBCP which becomes more pronounced when increasing separation of the flakes with the bulk. We propose a minimal model with odd parity superconductivity, which follows the spirit of Refs.\cite{Guinea12,Roldan13}, but with an additional band having nodal induced superconductivity. We calculate the tunneling current and find that the tip-sample interaction can be non-
perturbative, leading to a ZBCP formed by ABS induced by the interaction with the tip.

\begin{figure}[t]
\begin{center}
\includegraphics[width=0.7\textwidth]{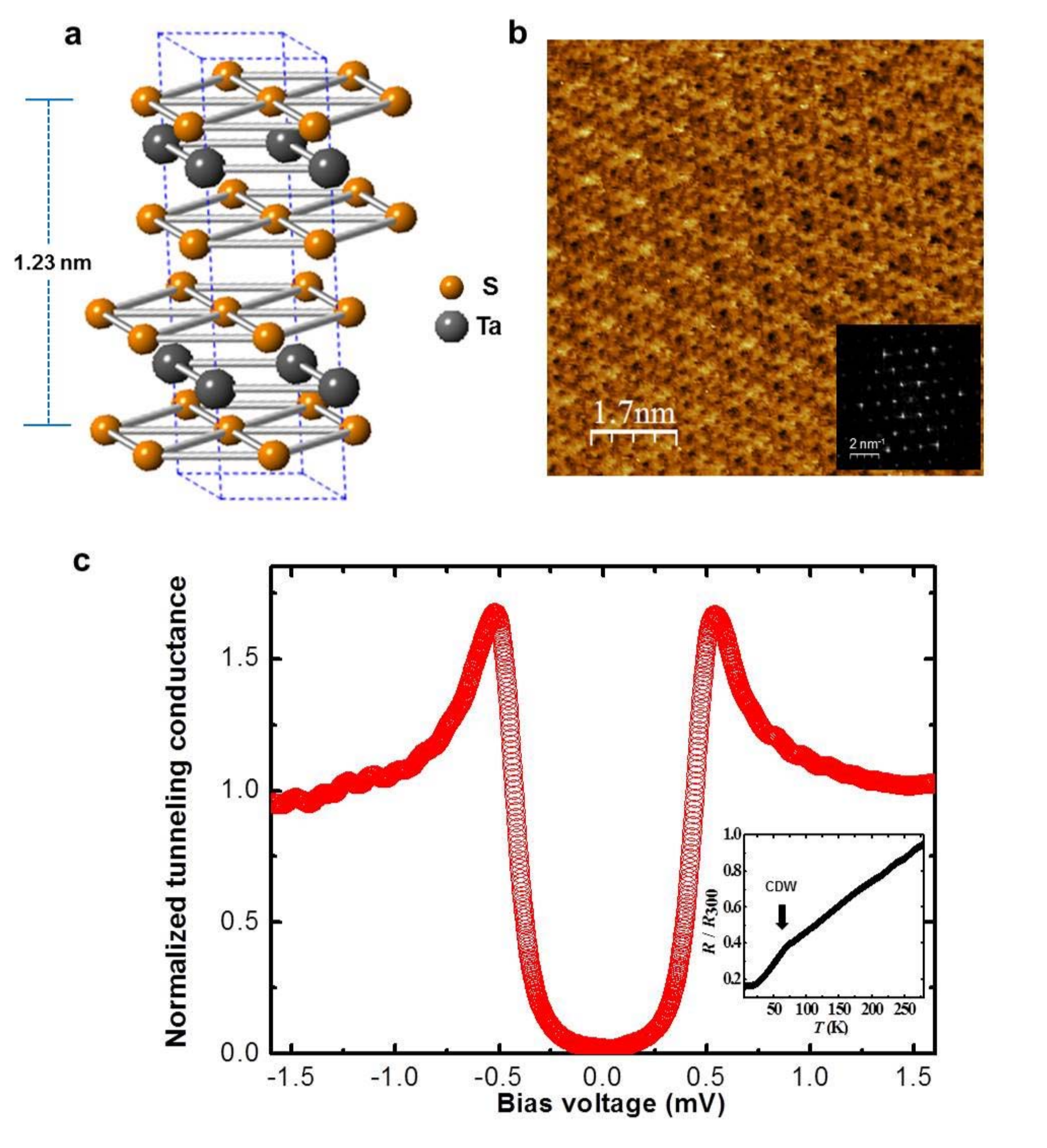}
\caption{(Color online). In a) we show the structure of 2H-TaS$_2$, and highlight the unit cell size. b) is an atomic resolution surface topography showing the characteristic features of large and flat surfaces of 2H-TaS$_2$ and c) shows the tunneling conductance of 2H-TaS$_2$ in the superconducting phase at 150 mK (inset shows the temperature dependence of the resistivity). These results are obtained on large flat areas of 2H-TaS$_2$,  and following figures instead in areas showing detached and broken flakes.}
\end{center}
\end{figure}

\section{Experimental.}

We measure a single crystalline sample of 2H-TaS$_2$ grown by chemical transport\cite{Gruehn00,Coronado10}. We use STM set-up described  elsewhere \cite{Suderow11} featuring an in-situ sample positioning system that allows to easily change the scanning window and cover macroscopically different portions of the sample. We take images with a tunneling conductance around 0.3 $\mu$S, unless stated otherwise, and a bias set-point of 2.5 mV. Samples have a lateral size of 5 mm and present a residual resistance ratio around 10 with a critical temperature of $T_c=3~{\rm K}$. The resistance vs. temperature shows a feature at 75 K due to the charge density wave (CDW) transition in this compound (inset Fig.\ 1c) \cite{Rossnagel11,Barnett-PRL2006,Rossnagel07,Ge-PRB2012}. Before cooling, we exfoliate the samples using the scotch tape method. We find some areas which are shiny and flat, and others with a large amount of detached and broken flakes. In the large and flat areas, superconductivity shows s-wave BCS gap features (Fig.\ 1)~\cite{Guillamon11,Wezel12}. In the other areas, we often identify pronounced steps, as the one shown in Fig.\ 2a, and places with fractured layers, as shown in Fig.\ 3a. Fractures and pronounced steps evidence detached flakes and are produced by exfoliation. Generally, scanning is more difficult on those areas. However, by searching many different scanning windows, we are able to obtain atomic resolution over some flakes.

\section{Results.}

We find that on the areas with detached flakes, the detached upper surface flakes show a ZBCP, whereas the lower surface a superconducting gap (Fig.\ 2b).  The ZBCP disappears at the $T_c$ of the bulk (Fig.\ 2c and d). The phenomenology is varied, with peaks of different sizes. For instance, in Fig.\ 3, the upper layer shows a fracture separating two regions with different tunneling conductance curves. The top left layer is a few tenths of nm more separated from the bottom layer than the top right layer (see profile in Fig.\ 3b). The tunneling spectroscopy on the bottom layer shows a superconducting gap, and the one on the top layers a ZBCP. On the top left layer the ZBCP is more pronounced than on the top right layer. There appears to be a correlation between the coupling of the detached layer and the size of the ZBCP.

\begin{figure}[t]
\begin{center}
\includegraphics[width=0.9\textwidth]{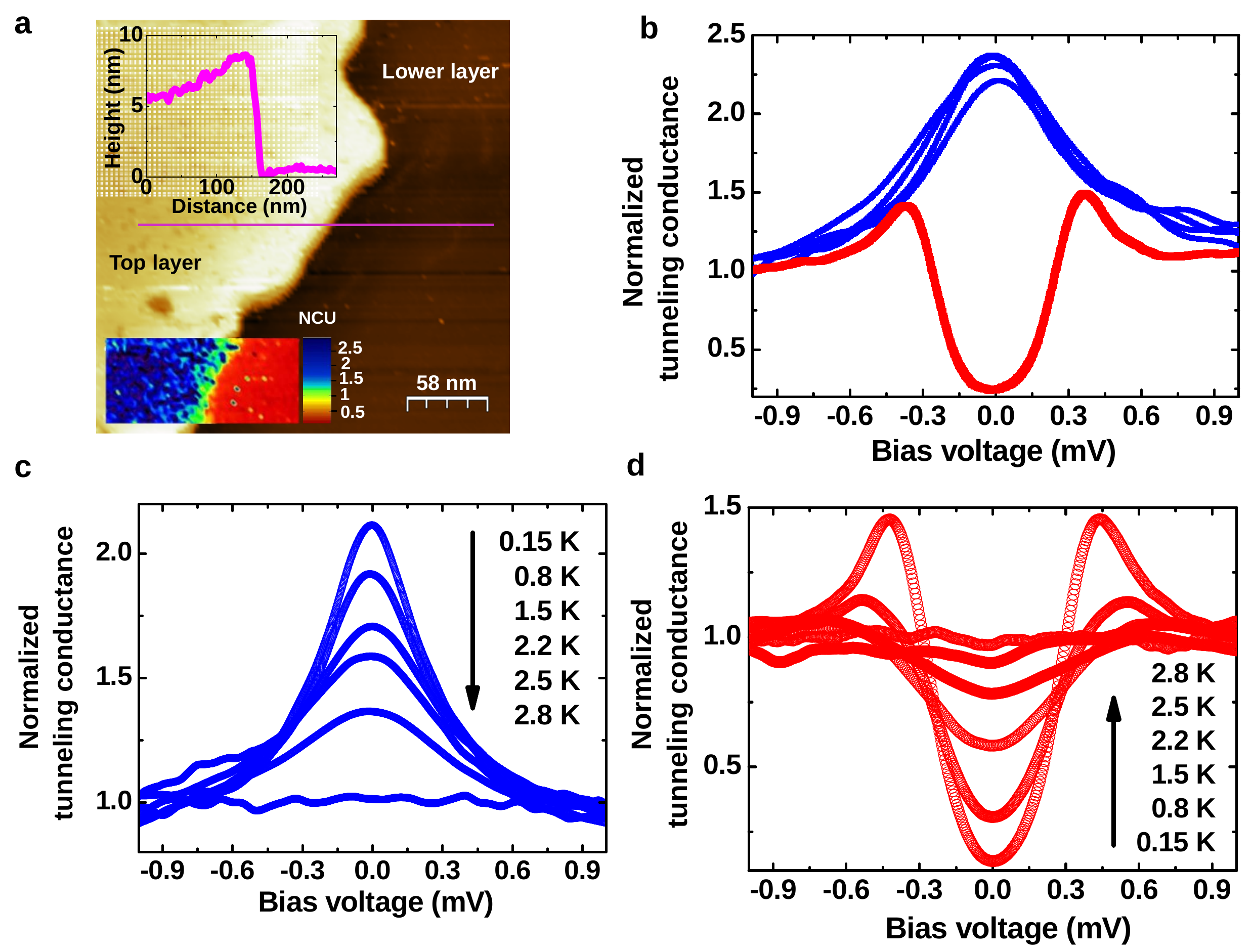}
\vskip -0.1 cm
\caption{(Color online). In (a) we show a topography of a surface of the sample showing a detached layer. The upper inset shows a profile through the step at the border of the layer, and the lower inset an image of the zero bias conductance (color scale is shown in units of conductance normalized at high bias voltages) taken at 0.15 K. The image has been pasted at the position where it was taken. In (b) we show tunneling conductance curves obtained throughout the top layer (blue) and on the bottom layer (red), both at 0.15 K. The corresponding temperature dependencies are shown in (c) and (d). \label{Fig2}}
\end{center}
\end{figure}

\begin{figure}[t]
\begin{center}
\includegraphics[width=0.9\textwidth]{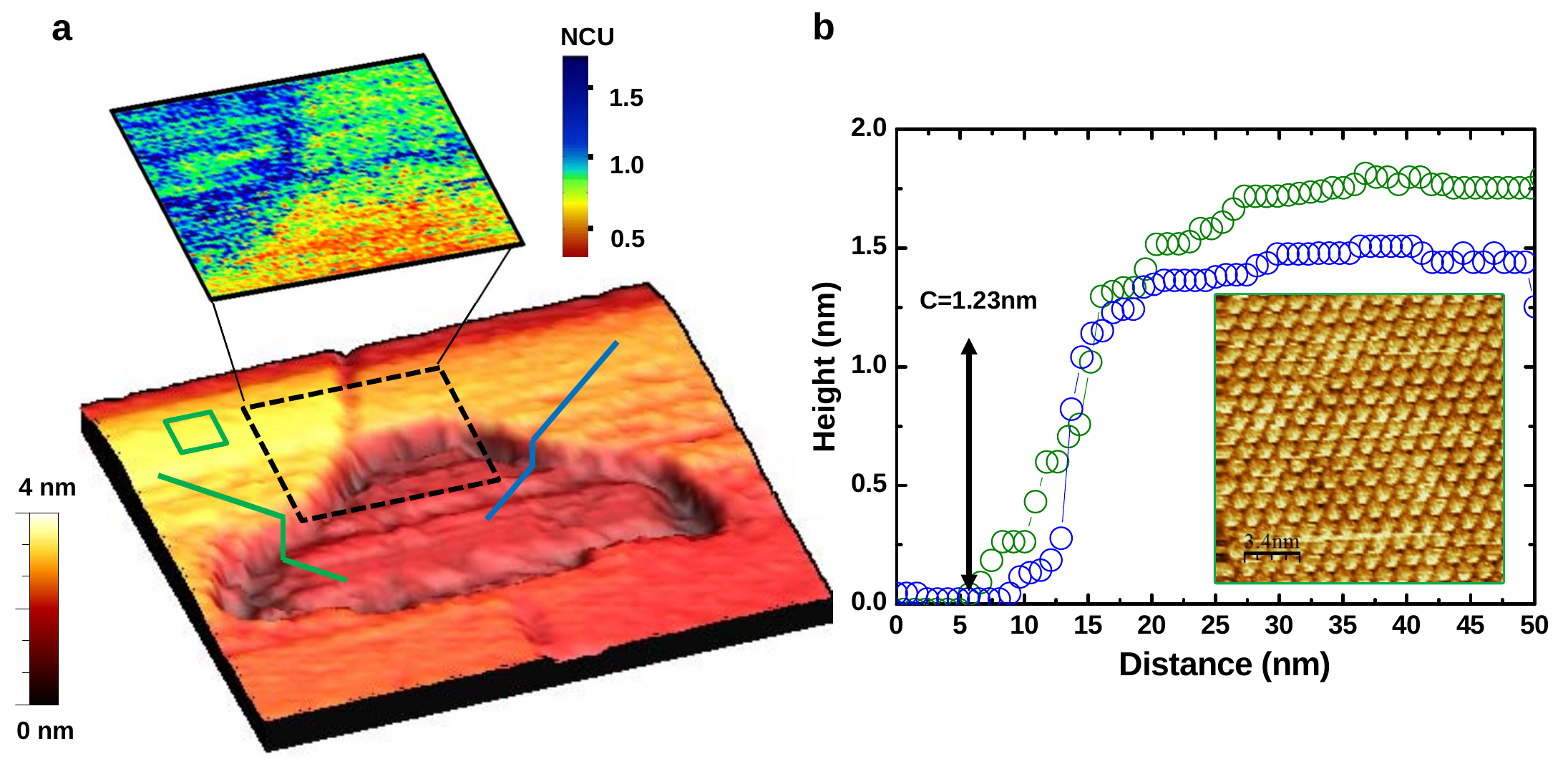}
\vskip -0.1 cm
\caption{(Color online) In (a) we show a topography (190 nm $\times$ 175 nm) of a region with an upper layer slightly separated from the lower layer. An image of the zero bias conductance on the area marked by the black square is also shown. In (b) we show profiles across two different directions, as marked by the green and blue lines of the image in (a). We give the unit cell size as an arrow in (b). In the inset we show an atomic scale topography image obtained in the area marked by a green square in a. Data are taken at 0.15 K.}
\end{center}
\end{figure}

\section{Modelling superconductivity in unit cell layers of {2H-TaS}$_2$.}

\subsection{Superconductivity in flakes.}

To analyze our results in 2H-TaS$_2$, we construct a minimal two-dimensional tight-binding model intended to reproduce main features of the Fermi surface of 2H-TaS$_2$ unit cell layers (Fig.\ 4a). We take a central nearly circular Fermi surface at the $\Gamma$ point and also nearly circular pockets at $K$ and $K'$ points. Electronic states on the Fermi surface have a strong $5d$ character, favoring an active role of electron-electron interaction as a leading superconducting instability \cite{KohnLuttinger1965}. The contribution of 
electron-electron interaction to $K-K'$ intervalley scattering is due to the short range part of the Coulomb potential. The latter consists in the Hubbard onsite repulsion between electrons with opposite spin in the same atomic orbital. This is not negligible for $5d$ electrons and favors a triplet odd-momentum pairing state, with gaps of opposite signs in different $K$ and $K'$ valleys,  
$\Delta_K=-\Delta_{K'}$ which breaks parity \cite{Guinea12,Mazin2010,Taniguchi12,Ye12,Roldan13}. In the even-parity $\Gamma$-centered Fermi pocket, superconductivity is induced with a nodal 
character.

To construct the wavefunctions, we consider a triangular lattice with two $d$-orbitals per site, $d_{x^2-y^2}$ and $d_{xy}$, up to nearest-neighbor hopping. The triangular lattice is specified by two basis vectors ${\bf a}_1=(a,0)$ and ${\bf a}_2=(a/2,\sqrt{3}a/2)$, and we also define 
${\bf a}_3={\bf a}_2-{\bf a_1}$. On the same site the two orbitals are orthogonal and degenerate. We  introduce Fermionic operators 
$c_{{\bf R},\alpha,\sigma}$ for electrons at site ${\bf R}$, with spin  $\sigma=\uparrow,\downarrow$ in orbital $\alpha=d_{x^2-y^2},d_{xy}$ 
and group them in the orbital vector ${\bf c}_{{\bf R},\sigma}$. A gap that changes sign between the $K$ and $K'$ valley can be 
modeled as an imaginary nearest neighbor pairing, in the spirit of the Haldane model \cite{Haldane1988}. The Hamiltonian can be compactly 
written as 
\begin{eqnarray}\label{Htot}
H&=&-\sum_{{\bf R},\sigma}\sum_{i=1}^3{\bf c}^{\dag}_{{\bf R},\sigma} h_i{\bf c}_{{\bf R}+{\bf a}_i,\sigma}+{\rm H.c.}
-i\delta \sum_{{\bf R}\alpha}\sum_{s=\pm,j=1}^3 c^{\dag}_{{\bf R}\alpha\uparrow}c^{\dag}_{{\bf R}+s{\bf a}_j,\alpha\downarrow}s(-1)^j
+{\rm H.c.},
\end{eqnarray}

where $(h_i)_{\alpha,\beta}\equiv h_{{\bf R},\alpha;{\bf R}+{\bf a}_i,\beta}$ for $i=1,2,3$ is the hopping matrix element between orbital 
$\alpha$ in site ${\bf R}$ and orbital $\beta$ in site ${\bf R}+{\bf a}_i$. The matrices $h_i$ are parametrized by two independent hoppings 
$t_{dd\sigma}$ and $t_{dd\pi}$ as in Ref.~\cite{Guillamon08}. In the Nambu basis defined by the 8-component vector 
$\boldsymbol{\psi}_{\bf k}=({\bf c}_{\bf k}, {\cal T}{\bf c}_{\bf k})^T=({\bf c}_{{\bf k},\uparrow},{\bf c}_{{\bf k},\downarrow},{\bf c}^{\dag}_{-{\bf k},
\downarrow},-{\bf c}^\dag_{-{\bf k},\uparrow})^T$, 
with ${\cal T}=is_y K$ the time reversal operator, the Bogoliubov - de Gennes Hamiltonian reads ${\cal H}_{\bf k}^{\rm BdG}=({\cal H}^0_{\bf k}-\mu)\tau_z+\Delta_{\bf k}s_z\tau_x$, 
where $s_i$ and $\tau_i$ are Pauli matrices acting on the spin and particle-hole subspaces, respectively, $\mu$ the chemical potential, 
${\cal H}^0_{\bf k}=-2\sum_{i=1}^3h_i\cos({\bf k}\cdot{\bf a}_i)$, and the triplet $f$-wave gap function is 
$\Delta_{\bf k}=2\delta\sum_{i=1}^3(-1)^i\sin({\bf k}\cdot{\bf a}_i)$. ${\cal H}^0_{\bf k}$ yields two spin-degenerate bands 
$\epsilon^{\pm}_{\bf k}$. Choosing the Fermi energy at half the bandwidth, $\mu=0$, for $t_{dd\pi}\leq t_{dd\sigma}/3$ the Fermi surface 
(Fig.~4 left) consists of two inequivalent pockets centered at the $K$ and $K'$ points, defined by the $\epsilon^-_{\bf k}=\mu$, and a 
pocket centered about the $\Gamma$ point, defined by $\epsilon^+_{\bf k}=\mu$. The odd-momentum gap function has strength 
approximately $\Delta_K\sim 4\delta$ on the $K$ and $K'$ valleys, and it induces a weaker pairing $\Delta_\Gamma\sim\delta$ on the 
$\Gamma$ pocket characterized by sign changing nodes  (see Fig.~4). 

Scattering at localized centers has been extensively considered by literature in nodal $d$-wave superconductors 
\cite{LeePRL1993,BalatskyPRL1993,BalatskyPRB1995,Balatsky06} and in graphene 
\cite{RyuPRL2001,Pereira-PRL2006,Peres-PRL2006}. Both share a particle-hole symmetric spectrum and a low energy Dirac-like dispersion.  
In nodal $d$-wave superconductors, scattering induced mixing of wave functions characterized by sign changing gap produces ABS, breaks 
pairs and leads to a decrease in $T_c$. Within our model, the leading superconducting instability, arising upon short range repulsive electron-
electron $K-K'$ intervalley scattering, generates a gap with strength $\Delta_K$ on the $K$ valley. But close to the Fermi energy the relevant energy 
scale is the reduced gap  $\delta$ on the $\Gamma$ pocket that determines the cutoff scale for the linear approximation around the 
nodal points.

\subsection{Tunneling into flakes with induced nodal superconductivity.}

Weak scattering in the $\Gamma$ pocket is enough to produce ABS because $\delta$ is small. We calculate the tunneling current using 
Nambu-Keldysh non-equilibrium formalism \cite{MartinRodero-PRL1994,LevyYeyati-PRB1995,CML96}, where the tunneling is treated at all orders 
in perturbation theory. Assuming a point-like tunneling of amplitude $t_0$, the current is determined by the 
momentum-averaged superconductor Green's function $\hat{g}_{\rm s}(z)=\sum_{\bf k}(z-{\cal H}^{\rm BdG}_{\bf k})^{-1}$, which, due to oddness of the gap function, yields no contribution from Andreev reflection in the tip. Following Ref.~\cite{CML96}, the conductance is
\begin{equation}\label{Transmission}
\sigma(\omega)=\frac{e^2}{h}\frac{4\pi^2t^2_0\rho_{\rm tip}(\omega-eV)\rho_{\rm sc}(\omega)}{\left|1-t_0^2g_{\rm sc}(\omega)g_{\rm tip}(\omega-eV)\right|^2},
\end{equation}
where $g_\alpha(\omega)={\rm Tr}~\hat{g}_\alpha(\omega+i0^+)/2$ and $\rho_\alpha=-{\rm Im}~g_\alpha/\pi$, with $\alpha={\rm tip}, {\rm sc}$. 
In the denominator of Eq.~(\ref{Transmission}) we recognize the tip self-energy, $\Sigma_{\rm tip}=t_0^2g_{\rm tip}$. Close to the Fermi energy 
it is natural to assume $\Sigma_{\rm tip}=-i\pi t_0^2\rho_{\rm tip}$, with $\rho_{\rm tip}$ constant tip DOS at the Fermi energy. A 
resonant behavior is reached when the denominator of Eq.~(\ref{Transmission}) goes to zero, 
\begin{equation}\label{PolesG}
1-\Sigma_{\rm tip}g_{\rm sc}(\Omega)=0.
\end{equation}
These are the poles of the tip-modified superconductor Green's function at the tip position. The problem is analogous to the case of a strongly 
scattering scalar  impurity in a $d$-wave nodal superconductor, where a localized virtually bound state is induced 
\cite{Balatsky06,BalatskyPRL1993,BalatskyPRB1995}. In our case, the source $\Sigma_{\rm tip}$ is purely imaginary. Following Ref.~\cite{BalatskyPRB1995},  by approximating the nodal points as Dirac cones in the limit $|\omega|\ll \delta$, we find a purely 
imaginary resonance $\Omega=i\delta x$, with $x\log x=1/\lambda$ , where $\lambda=\frac{1}{\sqrt{3}}N_{\rm f}t_0^2\rho_{\rm tip}/t_{dd\sigma}$ is the effective coupling, and $N_{\rm f}=12$ the total number of flavors (six spin-degenerate Dirac cones). The tip locally breaks time-reversal symmetry and in the limit $\lambda\to\infty$ produces a resonance at zero energy characterized by a large broadening of order of $\delta$.

\begin{figure}[t]
\begin{center}
\includegraphics[width=0.7\textwidth]{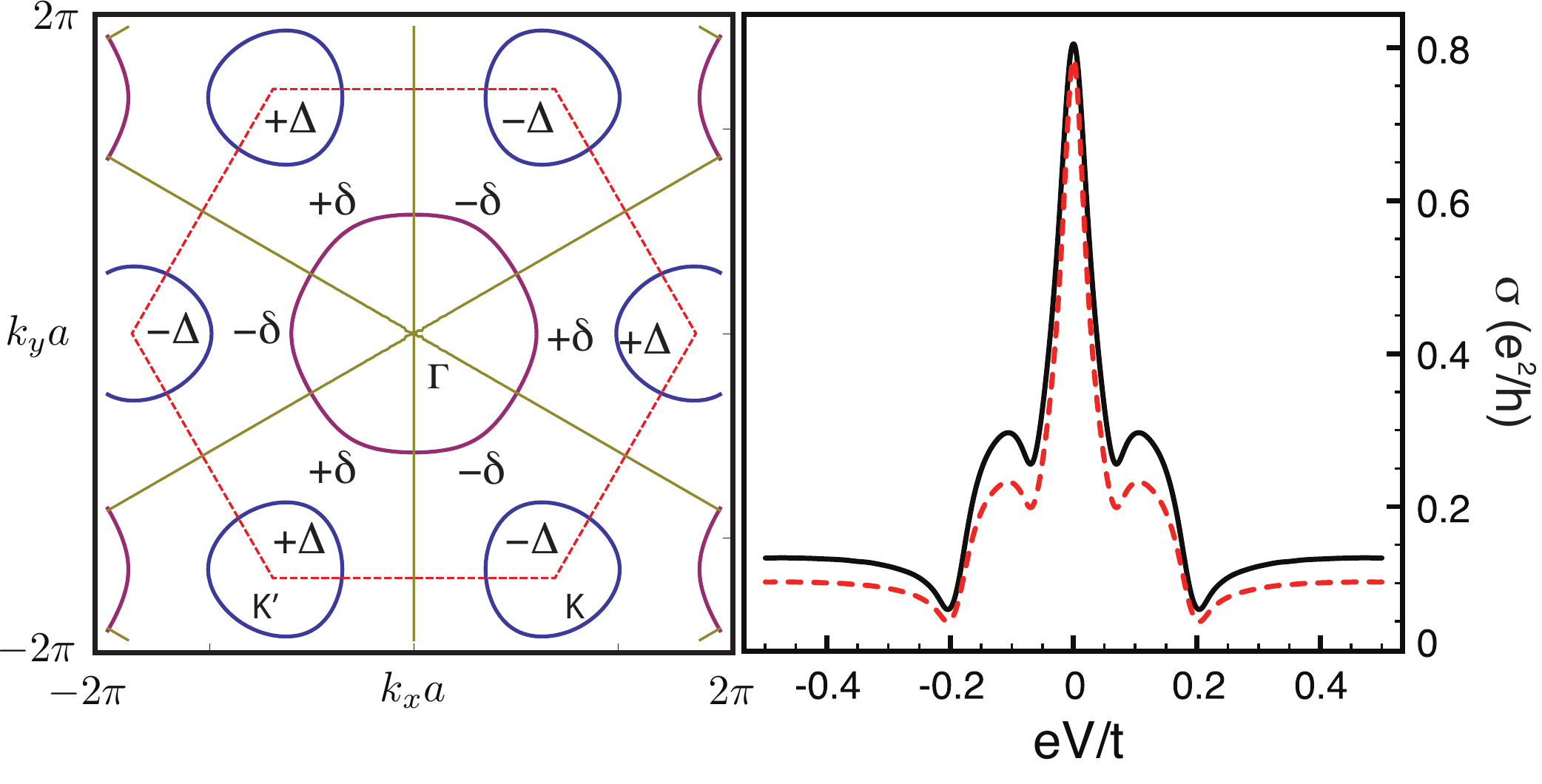}
\vskip -0.1 cm
\caption{(Color online) Left: Fermi surface of the TaS$_2$ obtained from the tight-binding model described in the text for the choice $t_{dd\sigma}=50~t_{dd\pi}\equiv t$. Zones of positive/negative gap and nodal lines are highlighted. On the $K,K'$ valleys (blue color) the gap is denote by $\Delta$ and on the $\Gamma$ pocket (violet color) by $\delta$. Red dashed line is the first Brillouin zone. Right: Conductance expected in a STM measurement as a function of the applied bias voltage $V$ for the two $d$-orbital tight-binding model. The gap is chosen to be $\delta/t_{dd\sigma}=0.05$. Coupling is non-perturbative with $\tilde\lambda=0.6, 0.8$ (red dashed and dark continous lines respectively, temperature of $\delta/10$). \label{Fig3}}
\end{center}
\end{figure}

We numerically calculate the tunneling conductance for the model 
Eq.~(\ref{Htot}) for a constant imaginary self-energy. We 
assume the tip couples to a particular site, with equal strength $t_0$ to the two $d_{x^2-y^2}$ and $d_{xy}$ orbitals. When dealing with the 
tight-binding model the coupling constant becomes $\tilde\lambda=t_0^2\rho_{\rm tip}/t_{dd\sigma}$, that is smaller than the effective low energy coupling constant $\lambda$ by about an order of magnitude, $\tilde\lambda=\lambda \sqrt{3}/(N_{\rm f})\ll \lambda$. A relatively weak tip-sample coupling can become non-perturbative at the low energy scale 
$\delta$. For very weak coupling $\tilde\lambda\ll1$ ($\lambda<1$) the conductance shows a local density of states with a gap at $\Delta_K$ on the $K$ and $K'$ Fermi pockets, and a small nodal gap $\Delta_\Gamma$ on the $\Gamma$ pocket. For larger coupling $\tilde\lambda\leq 1$ ($\lambda>1$) a broad ZBCP occurs (Fig.\ 4). The features indicative of the larger sized gap structure are washed out if coupling is unequal and stronger to the $\Gamma$ pocket. With large coupling, the interaction is non-perturbative and the 
size of the ZBCP, compared to the tunneling conductance, does not vary strongly.

The nodal superconductivity giving the ZBCP is weak because it is induced by the $K$ bands. Thus, the ZBCP appears for a range of relatively weak tip-sample 
couplings. Tip-sample coupling $t_0$ on order of $0.1\Delta_K$ remains perturbative with respect to the $K$ band but is 
enough to produce a sizeable ABS in the $\Gamma$ pocket.

\section{Discussion.}

In our experiment, we observe a ZBCP centered at the Fermi level that is essentially featureless. The ZBCP remains of same order within a conductance range of 0.2 and 2 $\mu$S. This can be expected, since the small induced gap on the $\Gamma$ pocket makes the perturbative regime with respect to this band inaccessible for operational values of the tunneling conductance. Thus, tunneling into the $\Gamma$ band remains non-perturbative and gives a ZBCP, whose size, as discussed below, depends on the coupling between the detached surface layer and the layer below.

2H-TaS$_2$ is a quasi-two dimensional layered transition-metal dichalcogenide and belongs to the series of 2H-MX$_2$ with $M={\rm Ta}, 
{\rm Nb}$ and $X={\rm S}, {\rm Se}$. Experiments report 
two-band superconducting gaps in 2H-NbSe$_2$ and in 2H-NbS$_2$~\cite{Hess90,Guillamon08PRB,Iavarone08,Kacmarcik10}. In 2H-TaSe$_2$, mixed 
surfaces are observed, which consist of a unit cell superconducting 2H-TaSe$_2$ over non-superconducting 1T-TaSe$_2$ crystallizing in 
a trigonal structural polytype. These superconducting unit cell 2H-TaSe$_2$ layers give also images with a ZBCP in the tunneling 
conductance\cite{Galvis13}. The model presented here can also explain qualitatively this behavior. There are some relevant differences, 
however. In 2H-TaS$_2$ the ZBCP is more difficult to find than in 2H-TaSe$_2$ and appears only by intensive searching for detached 
layers. After exfoliation with scotch tape, 2H-TaSe$_2$ has a stronger tendency to form large flat areas with mixed polytypes close to the surface, and 2H-TaS$_2$ shows 
often nice flat surfaces, separated by areas with detached layers. Moreover, strong atomic scale modulations are observed in the size of the ZBCP in 2H-TaSe$_2$. By contrast, the ZBCP discussed here in 2H-TaS$_2$ is roughly homogeneous at atomic scale. 
The van der Waals gap between $MX_2$ layers increases from 2H-NbS$_2$, 2H-NbSe$_2$, 2H-TaS$_2$ and 2H-TaSe$_2$, with the latter two compounds having a mostly two-dimensional Fermi surface \cite{Rossnagel11}. Stronger coupling between layers for the Nb compounds may explain why ZBCP have only been observed in the Ta compounds. The form and electronic properties of the interlayer coupling seem rather important in the formation of ZBCP.

Our experiment and model suggests that there can be an interesting crossover behavior from a nodal odd-momentum gap in the detached flake case, that manifests itself with the appearance of a ZBCP, to an ordinary BCS gap in the bulk case, depending on the coupling between layers. For instance, the peak in Fig.\ 3 increases with the step height. We expect that coupling with the substrate affects the whole layer, and leads to 
averaging of sign changing gap features, and not localized Andreev levels as those formed by a point scatterer. At present, it appears challenging 
to better control interlayer coupling and study this effect quantitatively. But this may become possible in future using intercalation of molecular compounds between layers 
\cite{Coronado10}, and would open a way to modify parity and spatial structure of pairing interactions.

The T$_c$ values where the ZBCP disappears are of order of those found in the substrate. The T$_c$ in the transition metal dichalcogenides can vary between 0.1 K and 10 K \cite{Wilson75} depending on pressure or defects in the crystal. Thus, internal strain is probably important in explaining superconducting T$_c$ in layers. New superconducting properties might be unveiled by further improving unit cell layer synthesis and measurement.

\section{Conclusions.}

In summary, we provide a model according to which unit cell layers of two-dimensional 2H-TaS$_2$ present induced nodal superconductivity that can be perturbed by weak scattering. Extended Andreev excitations at the Fermi level can arise without a significant pair breaking effect on the main superconducting bands. The proposed mechanism suggests a way to generate ABS in a controlled way and can be extended to other non-conventional superconducting systems. Experiments in detached flakes of the transition metal dichalcogenide 2H-TaS$_2$ present a strong zero bias conductance peak, which can be understood as an ABS produced by nodal induced superconductivity, as proposed in our model. This suggests that flakes a few atoms thick could show very different superconducting properties than the bulk counterpart.

\section{Acknowledgments}
This work was supported by the Spanish MINECO (Consolider Ingenio Molecular Nanoscience CSD2007-00010 program and FIS2011-23488, ACI2009-0905, FIS2011-23713 and MAT2011-22785), by the European Union (ERC Advanced Grants 290846 and 247384; Graphene Flagship contract CNECT-ICT-604391, Marie Curie Actions FP7-PEOPLE-2013-CIG-618321 and COST MP1201 action), the Comunidad de Madrid through program Nanobiomagnet and by the Generalidad de Valencia through programs PROMETEO and ISIC-NANO. L. C. acknowledges discussions with F. Taddei. We acknowledge technical support of UAM's workshops, SEGAINVEX. 

%\bibliography{LastBib_noTitle}

\end{document}